\documentclass[12pt,twoside]{article}
\usepackage{fleqn,espcrc1}
\usepackage{psfig}

\newcommand{\AmS}{{\protect\the\textfont2
  A\kern-.1667em\lower.5ex\hbox{M}\kern-.125emS}}
\title{Measurements of r-process nuclei}

\author{
K.-L.~Kratz\address[KCHM]{Institut f\"ur Kernchemie, 
Universit\"at Mainz, D-55128 Mainz, Germany}
}

\begin{document}

% typeset front matter
\maketitle

\begin{abstract}
Progress in the astrophysical understanding of r-process nucleosynthesis 
also depends on the knowledge of nuclear-physics quantities of extremely 
neutron-rich isotopes. In this context, experiments at CERN-ISOLDE have 
played a pioneering role in exploring new shell-structure far from stability. 
Possible implications of new nuclear-data input on the reproduction of 
r-abundance observations are presented.   
\end{abstract}

\section{General motivation}

Relative to other fields, explosive nucleosynthesis is propably unique
in its requirements of a very large number of (nuclear-) physics quantities 
in order to achieve a satisfactory description of the astrophysical phenomena 
under study. Because nuclei of extreme N/Z composition exist in explosive 
environments, an understanding of their nuclear-structure properties has 
always been and still is a stimulating challenge to the experimental and 
theoretical cosmo-chemistry and nuclear-physics community. 
   
\section{R-abundance peaks and neutron shell-closures}

About 40 years ago, Charles D. Coryell, in his important but widely 
unrecognized article on {\it ''The Chemistry of Creation of the Heaviest 
Elements''} \cite{klk:Cor61}, has summarized his ideas about the importance 
of shell-structure from the 1950's for the formation of elements of mass 
number A$>$70 by rapid neutron capture. Starting with the realization, 
that historically {\it ''chemists have been interested from time immemorial 
in the chemical composition of the world around us''}, he emphasizes the 
importance of the neutron shell-closures at N=50, 82 and 126 for the 
time-scale of the fast neutron-capture process and the {\it ''r-process 
pile-up''} at the bromine, xenon and platinum peaks of Suess and Urey's 
so-called ''cosmic abundances'' \cite{klk:Suess56}. As {\it ''primary 
antecedent} (odd-Z) {\it species with N=50''}, already at that 
time Coryell favours $^{79}$Cu and $^{81}$Ga. He considers these isotopes 
out of experimental reach, because they are {\it ''even more neutron-rich 
than primary fission products''}. On the basis of their isobaric {\it 
''displacement from the valley of stability''}, he estimates a half-life 
of about 0.1~s. Similarly, for N=82, Coryell suggests that the decay of 
the precursors $^{127}$Rh, $^{129}$Ag and $^{131}$In forms the top of the 
A$\simeq$130 abundance peak; and for A$\simeq$195 peak {\it ''topped by 
$^{195}$Pt''} he assigns $^{195}$Tm as the primary N=126 nucleus. With his 
further assumption {\it ''that the energy of first decay, and thus the 
probable half-life for the decay, stays about constant''} along the 
r-process path, the total time required for a steady-state build-up of 
heavy nuclides from Fe to U would be roughly 6.5~s. It is interesting to 
compare this estimate with the contemporary one of Burbidge et al. 
\cite{klk:b2fh} of about 80~s.

\begin{figure}
%\centerline{\psfig{file=klbild9neu.eps,height=6.cm}}
\centerline{\psfig{file=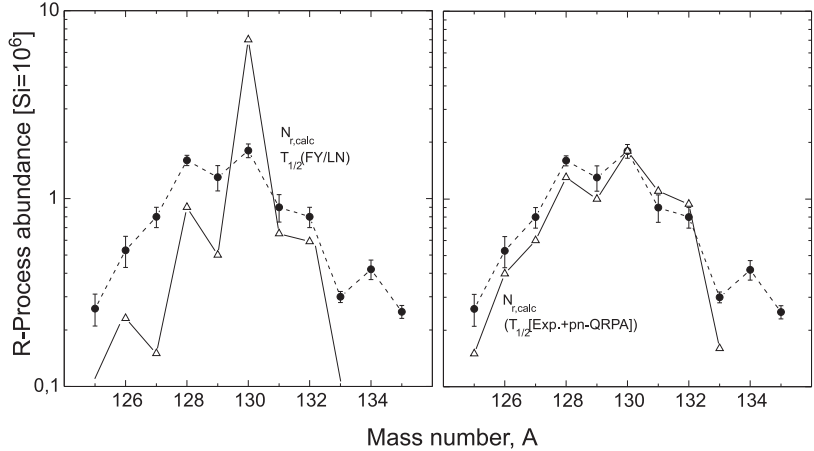,height=6.cm}}
%\vspace{6cm}
\caption{
Static calculations of A$\simeq$130 r-abundances assuming S$_n$=const and 
N$_{r,\odot}$(Z)$\times$${\lambda}_{beta}$(Z)=const. In the left part, 
theoretical T$_{1/2}$ and P$_n$ values from straight-forward QRPA 
calculations \protect{\cite{klk:MORA}} neglecting effects from the p-n 
residual interaction are depicted. The right part shows an improved fit 
with the same $\beta$-flow assumptions,  but now using measured 
$\beta$-decay data and QRPA predictions fine-tuned to reproduce 
nuclear-structure properties around $^{132}$Sn known in the early 1990's. 
In both cases, masses from an early version of the 1995 FRDM 
\protect{\cite{klk:FRDM}}
were used in the calculations of the gross $\beta$-decay properties. }
\label{klk1}
\end{figure}      

Today, we know that the r-process matter flow is not constant. It is fast 
in the mass regions between the magic neutron numbers, whereas the 
shell-closures act as ''bottle-necks''. Here, at N=50, 82 and 126, each, the 
classical r-process needs several hundred milliseconds to pass these 
regions, thereby building up the well known r-abundance peaks observed in 
the solar system (N$_{r,\odot}$). Therefore, the nuclear-physics input to 
these N$_{r,\odot}$ maxima is still today of particular importance for the 
understanding of r-process nucleosynthesis, independent of its stellar site. 
     
Still some 15 years ago, astrophysicists believed that nuclear-structure 
information on r-process nuclei, lying 14 ($_{28}$Ni) to 36 ($_{63}$Eu) 
mass units away from the $\beta$-stability line, would never be accessible 
to experiments on earth. 
However, continued progress in ion-source and mass-separator technology has 
soon after resulted in the identification of the first two classical,
neutron-magic ''waiting-point'' isotopes \cite{klk:b2fh}, $^{80}$Zn$_{50}$ 
and $^{130}$Cd$_{82}$ at OSIRIS, TRISTAN and ISOLDE, respectively 
\cite{klk:lund,klk:gill,klk:Kratz86}. As was shown in
\cite{klk:Kratz86,klk:Kratz88}, with their $\beta$-decay properties 
first evidence for the existence of the earlier postulated 
N$_{r,\odot}$(Z)$\times$$\lambda_\beta$(Z)$\approx$const. correlation 
could be achieved. We would, however, remind the r-process laymen 
that this is a rather crude correlation based on simple {\bf static} 
calculations with the assumption of constant neutron-separation energies 
(S$_n$) for all (N$_{mag}$+1) nuclei at the ''staircases'' associated with 
the rising sides of the N$_{r,\odot}$ peaks (see, e.g., Fig. V,3 in 
\cite{klk:b2fh}, or Figs. 9 and 10 in \cite{klk:Kratz88}). In this approach, 
exclusively the $\beta$-decay rate ($\lambda$$_{\beta}$) or the half-life 
(T$_{1/2}$=ln2/$\lambda$$_{\beta}$) of the neutron-magic nuclides are of 
importance. Nevertheless, at those times when the predictions for the 
T$_{1/2}$ of N=82 $^{130}$Cd, for example, varied between 30~ms and 1.2~s, 
our experimental value of T$_{1/2}$=(196$\pm$35)~ms was of considerable 
importance to constrain the equilibrium conditions of an r-process within 
the above approximation. The result of such calculations of the 
A$\simeq$130 N$_{r,\odot}$ peak some 10--15 years ago is depicted in  
Fig.~\ref{klk1}. In the left part, straightforward T$_{1/2}$ calculations 
are shown using the Quasi-Particle Random-Phase Approximation (QRPA) of 
Ref. \cite{klk:MORA} with Folded-Yukawa wave functions, a Lipkin-Nogami 
pairing model and the (presumably too low) masses from an early version of 
the 1995 Finite-Range Droplet Model (FRDM) \cite{klk:FRDM}, but neglecting 
proton-neutron (p-n) residual interaction. For comparison, in the right 
part of this figure, our again static r-abundance calculations from the 
early 1990's with experimental data on $^{131,133}$In and $^{130}$Cd and 
an improved QRPA version, now taking into account effects of the p-n 
interaction are depicted (see, e.g. Ref. \cite{klk:krfr}). Nowadays, with 
our largely improved knowledge on nuclear masses and $\beta$-decay 
properties of nuclei far from stability as input for time-dependent 
''canonical'' or dynamical r-process network calculations (see, e.g., 
\cite{klk:Kratz93,klk:fkt94,klk:11,klk:16}), the above simplistic static 
picture should therefore no longer be used for abundance simulations. 

In any case, the experimental success in the late 1980's strongly motivated 
further experimental and theoretical nuclear-structure investigations and 
its possible astrophysical implications. For example, the second N=50 
waiting-point isotope $^{79}$Cu could be identified \cite{klk:Kratz91}. 
Furthermore, the measured high delayed-neutron emission probabilities 
(P$_n$) of odd-Z r-process nuclei just beyond N=50 were shown 
to be the nuclear-structure origin of the odd-even staggering in the 
N$_{r,\odot}$ peak at A$\simeq$80 \cite{klk:Kratz90}; and from the 
interpretation of the $^{80}$Zn$_{50}$ decay scheme, first evidence for 
a vanishing of the spherical N=50 shell closure far from stability was 
obtained \cite{klk:KratzPRC88}. 
 
Since the late 1980's, important progress has been achieved in the 
understanding of the systematics and development of nuclear-structure 
with increasing neutron excess (see, e.g. 
\cite{klk:BadHoneff,klk:NFFS,klk:ENAM96}), with 
CERN-ISOLDE always playing a leading role in this field. However, due to 
the generally very low production yields, the restriction to chemically 
non-selective ionization modes, or the application of non-selective 
detection methods, no further information on isotopes lying {\bf on} the 
r-process path could be obtained for quite some time. Only in recent years, 
the identification of additional r-process nuclides has become possible 
thanks to considerable improvements of the {\bf selectivity} in production 
and detection methods of rare isotopes, e.g. by applying resonance-ionisation 
laser ion-source (RILIS) systems.            

\section{Recent experiments}

Today, there are mainly three mass regions, where nuclear-structure 
properties of specific interest to r-process nucleosynthesis can be 
studied experimentally. The first is the classical r-process seed region 
involving very neutron-rich Fe-group isotopes in the vicinity 
of the doubly-(semi-)magic nuclei $^{68}$Ni$_{40}$ and $^{78}$Ni$_{50}$. 
Recent spectroscopic results can, for example, be found in 
\cite{klk:ENAM98,klk:ISP}. The second region is that of far-unstable nuclei 
around A$\simeq$115. Here, most canonical r-process calculations show a 
pronounced r-abundance trough, which we believe to be due to nuclear-model 
deficiencies in predicting ground-state shapes and masses at and beyond 
N=72 mid-shell (see, e.g. \cite{klk:Kratz93,klk:10}). Recent experimental 
information on that mass range can be found, e.g. in 
\cite{klk:ENAM98,klk:ISP,klk:SAN,klk:GREN}. The third region of interest, 
on which the main focus of this paper has been put, is that around the 
doubly-magic nucleus $^{132}_{\phantom{1}50}$Sn$_{82}$. Apart from its 
astrophysical importance ({\it the r-process matter flow through the 
A$\simeq$130 N$_{r,\odot}$ peak}) \cite{klk:Kratz93,klk:11}, this area is 
of considerable shell-structure interest. The isotope $^{132}$Sn itself, 
together with the properties of the nearest-neighbour single-particle 
($^{133}_{\phantom{1}51}$Sb$_{82}$ and $^{133}_{\phantom{1}50}$Sn$_{83}$) 
and single-hole ($^{131}_{\phantom{1}50}$Sn$_{81}$ and 
$^{131}_{\phantom{1}49}$In$_{82}$) nuclides are essential for tests of the
shell model, and as input for future nuclear-structure calculations towards 
the neutron-drip line.

The bulk of data so far known in this mass region have been obtained from
$\beta$-decay spectroscopy at the mass-separator facilities OSIRIS and 
ISOLDE \cite{klk:NFFS,klk:ENAM96,klk:ENAM98,klk:ISP,klk:SAN,klk:GREN}. 
During the past few years, new data ''northeast'' of $^{132}$Sn were 
published also from spontaneous-fission studies of $^{248}$Cm (see, e.g., 
Korgul et al. \cite{klk:ko}, and Refs. therein). With respect to the  
closest neighbours of $^{132}$Sn, the structures of $^{131}$Sn$_{81}$ 
($\nu$-hole) and $^{133}_{\phantom{1}51}$Sb ($\pi$-particle) are fairly well 
known since more than a decade. More recently, the lowest $\nu$-particle 
states in $^{133}$Sn$_{83}$ have been identified at the General Purpose 
Separator (GPS) of the PS-Booster ISOLDE facility (see, e.g., Hoff et al. 
\cite{klk:Hoff}). From these data, valuable information on the spin-orbit 
splitting of the 2f- and the 3p-orbitals was obtained. These results were 
compared to mean-field and HFB predictions, and it was found that none of 
the potentials used in the past in {\it ab inito} shell-structure 
calculations was capable of properly reproducing the ordering and spacing 
of these states.  

With these new data, the close $^{132}$Sn valence-nucleon region is nearly 
complete. The only missing information are the $\pi$-hole states in 
$^{131}_{\phantom{1}49}$In, which can in principle  be studied through 
$\beta$-decay of the exotic nucleus $^{131}$Cd$_{83}$. Off-line test 
experiments to find a well-suited excitation scheme for resonance ionization 
of Cd were initially carried out in Mainz \cite{klk:92}; and first results 
on T$_{1/2}$ data and delayed-neutron emission probabilities (P$_n$) of N=82 
to 84 $^{130-132}$Cd using a similar laser system were obtained at ISOLDE. 
A detailed discussion of the initially quite surprising results for N=83 
$^{131}$Cd and N=84 $^{132}$Cd can be found in Ref. \cite{klk:Han00}. 
Already these gross properties indicate again that the structure below and 
beyond $^{132}$Sn is not at all well understood by now. Of direct 
astrophysical importance is our new measurement of the N=82 
waiting-point nucleus $^{130}$Cd. Its considerably improved half-life of 
T$_{1/2}$=(167$\pm$7) ms is somewhat lower than our old value from 1986 of 
(195$\pm$35) ms \cite{klk:Kratz86}, but still lies within the error limits 
given at that time. The measured P$_n$ value of (3.5$\pm$1.0) $\%$ is in 
agreement with our earlier estimate.  

Another important observation with respect to the calculation of $\beta$-decay 
properties of so far unknown N$\simeq$82 r-process waiting-point isotopes 
is that model predictions of Q$_\beta$ values in the region ''south'' of 
$^{132}$Sn differ considerably, with the general tendency that those models 
which exhibit a strong N=82 shell closure (e.g., FRDM \cite{klk:FRDM} or 
ETFSI-1 \cite{klk:ET1}), give the ''lower'' values. Recent increasing 
experimental evidence of a weakening of the classical shell-structure below 
$^{132}$Sn (see, e.g. 
\cite{klk:ENAM98,klk:ISP,klk:Han00,klk:NUBASE,klk:Kau00}),
however, seem to favour the systematically higher Q$_\beta$ predictions from 
mass models with ''shell-quenching'' (e.g., HFB/SkP\cite{klk:HFB} or 
ETFSI-Q\cite{klk:ETQ}) below $^{132}$Sn. 
This trend is of particular importance for the T$_{1/2}$ and P$_n$ 
predictions of Z$<$50, N=82 r-process waiting-point nuclei. Sizeable 
differences in the Q$_{\beta}$ values between un-quenched and quenched mass 
models start at $^{131}$In with $\Delta$Q$_{\beta}$$\simeq$400~keV, continue 
via $^{130}$Cd to $^{127}$Rh with about 1 MeV, and range down to $^{122}$Zr 
with $\Delta$Q$_{\beta}$$\simeq$4~MeV, resulting in considerable changes 
in theoretical T$_{1/2}$ for these isotopes. For $^{130}$Cd, for example, 
the T$_{1/2}$(QRPA) with Q$_{\beta}$(ETFSI-Q)=8.30~MeV is a factor 2.5 
shorter than with Q$_{\beta}$(FRDM)=7.43~MeV. 
It is interesting to note in this context that the same trend is competely 
independently predicted also by very recent shell-model calculations of 
B.A. Brown et al. \cite{klk:ab}. Therefore, in order to solve the above 
discrepancies, high-precision mass measurements in the region below 
$^{132}$Sn are foreseen at ISOLDE using the Penning-trap mass spectrometer 
ISOLTRAP. In any case, in our more recent r-process calculations, we prefer 
to use newly calculated T$_{1/2}$(GT+ff) values, which supercede our 1996 
evaluation \cite{klk:kpm} applied in earlier r-process studies. It should 
be pointed out in this context, that already since the early 1990's our 
astrophysics collaboration uses steadily updated evaluations of experimental 
and theoretical T$_{1/2}$ and P$_n$ values, in which the calculated gross 
properties are based on the FRDM mass and deformation parameters 
\cite{klk:FRDM} and contain a microscopic GT-strength \cite{klk:MORA} and 
a schematic ff-strength \cite{klk:tak}. For neutron-rich A$\simeq$130 
isotopes, this led already in the past to shorter T$_{1/2}$ and smaller 
P$_n$ values than listed in \cite{klk:MNK}. Therefore, we would appreciate 
if the rather general statement in a number of recent papers that some new 
shell-model half-lives are significantly shorter than those {\it ''currently 
adopted in r-process simulations''} could be put more precisely in the future. 
At least, as far as our Basel--Los Alamos--Mainz collaboration is concerned, 
our theoretical T$_{1/2}$(GT+ff) are quite similar to the new shell-model 
values. Hence, statements that those new values would speed up the r-matter 
flow in the neutron-magic bottle-neck regions do not concern our calculations. 

Another recent study of nuclear-structure developments towards N=82 is 
related to the $\beta$dn- and $\gamma$-spectroscopic measurements of 
neutron-rich Ag nuclides at CERN-ISOLDE, again using an optimized RILIS 
system for this element. This Z-selectivity was of considerable assistance 
in minimizing the activities from unavoidable surface-ionized In and Cs 
isobars. In this context, the additional selectivity of the spin- and 
moment-dependent hyperfine (HF) splitting was used to enhance the ionization 
of either the $\pi$p$_{1/2}$ isomer or the $\pi$g$_{9/2}$ ground-state 
(g.s.) decay of the Ag isotopes. Details about first $\beta$dn- and 
$\gamma$-spectroscopic applications of this technique to short-lived isomers 
of $^{122}$Ag up to the classical N=82 r-process waiting-point nuclide 
$^{129}$Ag can be found, e.g. in \cite{klk:SAN,klk:ENAM98,klk:ISP}. Because 
of its importance for r-process calculations, in particular for the time 
spent in the A$\simeq$130 N$_{r,\odot}$ peak region, some details about the 
difficulties in the measurement of this latter isotope are given. Only after 
fine-tuning the lasers to an off-center frequency leading to an enhancement 
of the ionization of the $\pi$g$_{9/2}$ level relative to the 
$\pi$p$_{1/2}$ state, an unambiguous identification of the g.s. 
$\beta$dn-decay of $^{129}$Ag with a T$_{1/2}$=(46$^{+5}_{-9}$) ms was 
possible (see, e.g. Fig.~10 in \cite{klk:ISP}). 
The measured T$_{1/2}$(g.s.) value 
is longer than our initial QRPA(Nilsson/BCS) estimate of 15~ms, but is in 
good agreement with our more recent QRPA(FY/LN) prediction \cite{klk:MNK} 
and two subsequent shell-model calculations of Refs. \cite{klk:mp,klk:ab}. 
However, the measured T$_{1/2}$(g.s.) is lower than our old static 
waiting-point {\it requirement} of about 130~ms 
\cite{klk:Kratz88,klk:Kratz93}. Within this simple approach, the 46~ms would 
not be enough to build-up the $^{129}$Xe r-abundance to its solar value 
(see the left part of Fig.~\ref{klk1} in this context), but -- as we will 
see later (in Figs.~\ref{klk3},\ref{klk4}) -- for more realistic dynamic 
calculations with (non-constant) S$_n$ values, e.g. from the ETFSI-Q mass 
model, the measured T$_{1/2}$(g.s.) for $^{129}$Ag fits much better. 
Furthermore, when requiring a really ''perfect'' fit at A=129 one will have 
to consider also smaller effects from non-equilibrium phases of the r-process, 
e.g. neutron capture and neutrino processing during freeze-out. Under such 
conditions, also the T$_{1/2}$ contribution from the $\pi$p$_{1/2}$ isomer 
might be of importance for the  residual ''stellar'' half-life of $^{129}$Ag. 
We calculate a value of T$_{1/2}$(GT+ff)$\simeq$125 ms for this isomer, in 
reasonable agreement with the very recent shell-model prediction of 89~ms by 
B.A. Brown. As is discussed e.g. in \cite{klk:CGS}, with an isomeric half-life 
in this range, a stellar T$_{1/2}$$\simeq$60--72~ms would result after 
neutron capture on $^{128}$Ag to form an excited $^{129}$Ag nucleus, which 
then $\gamma$-cascades down to the isomer and the g.s. And, indeed, 
a careful comparison of the A=129 $\beta$dn-decay curves taken under 
''laser-off'' and ''laser-on'' conditions at central frequency, gave a 
first indication of a weak, longer-lived $^{129}$Ag component with a 
T$_{1/2}$ in the range 80--160~ms. Now, it will be important to use 
isomer-specific ionization in combination with isobar separation at the 
High-Resolution Separator (HRS) of ISOLDE to ascertain the existence of 
this $\pi$p$_{1/2}$ isomer with a half-life different from the g.s.-decay. 

Because of space limitations, our $\gamma$-spectroscopic data on the decay 
of heavy Ag isotopes, in particular the extension of the 2$^+$ and 4$^+$ 
level systematics up to $^{128}$Cd and tentatively to N=82 $^{130}$Cd, 
cannot be discussed here. Therefore, we only mention that our data indicate 
a weakening of the spherical N=82 shell below doubly-magic $^{132}$Sn, as 
predicted by some theories. For details, we refer to our recent publications 
\cite{klk:ISP,klk:Kau00}. Finally, also our very recent first experiment on 
the decays of very neutron-rich Sn isotopes, again using RILIS at ISOLDE, 
should be mentioned \cite{klk:she}. The data are not yet fully analyzed, 
but already now the new $\beta$-decay properties of $^{135-137}$Sn indicate 
that they may help to improve our r-abundance fits in the A=134 to 
A=140 mass region (see Figs.~\ref{klk3},\ref{klk4}). 

\section{Fitting the A$\simeq$130 N$_{r,\odot}$ peak region} 

In particular since the recent astronomical observations of the possible 
existence of (at least) two r-process components from different sites, a 
{\bf primary main} r-process responsible for the heavy elements between Cd 
and the Th--U region \cite{klk:a50}, and a {\bf secondary weak} r-process 
requested to produce the lighter elements in and beyond the A$\simeq$80 peak 
\cite{klk:pfa,klk:tru}, the N=82, A$\simeq$130 waiting-point region has 
gained additional importance as the {\it first} bottle-neck in the 
main-component r-matter flow. Now -- without the N=50, A$\simeq$80 peak -- 
the Xe--Te peak determines to a considerable extent the total time needed 
for this process in the (still favourably discussed) SN~II scenario (see, 
e.g. \cite{klk:woo,klk:15,klk:16}). Based on our new experimental T$_{1/2}$ 
and P$_n$ values for $^{130}$Cd and $^{129}$Ag, together with the considerably 
improved understanding of the nuclear-structure properties in the $^{132}$Sn 
region which have been incorporated into our recent QRPA calculations for the 
so far unknown (Z$<$47) N=82 waiting-point isotopes, a quite satisfactory 
reproduction of the total r-abundance curve can be obtained in time-dependent, 
multi-component ''canonical'' r-process calculations (see, e.g. 
\cite{klk:CGS,klk:ISP,klk:16}). 

\begin{figure}
\centerline{\psfig{file=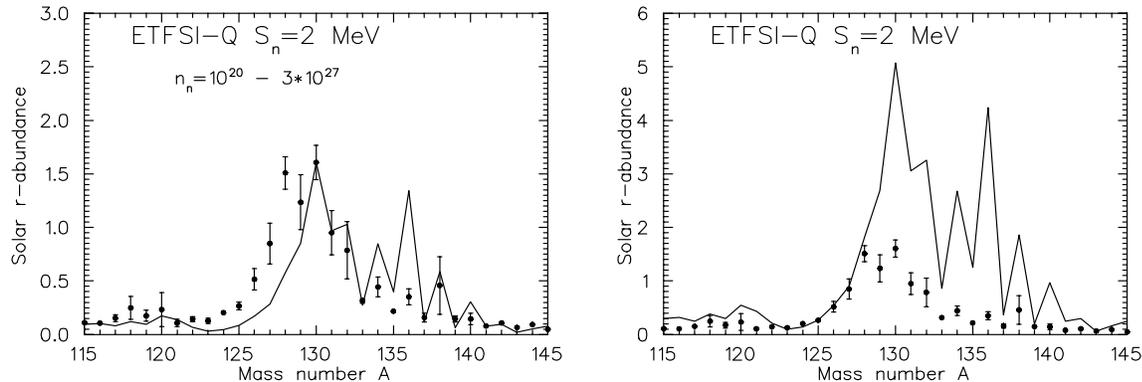,width=15.cm,angle=90}}
%\vspace{5cm}
\caption{R-abundances in the A$\simeq$130 region from dynamical calculations 
with the simplistic assumption of S$_n$=const for all N=83 isotones between 
$_{40}$Zr and $_{49}$In, in the left part ''normalized'' to the top of the 
N$_{r,\odot}$ peak at A=130, and -- alternatively -- in the right part to the 
bottom of the rising part of the peak at A=125.  
}
\label{klk3}
\end{figure}

\begin{figure}
%\vspace{5cm}
\centerline{\psfig{file=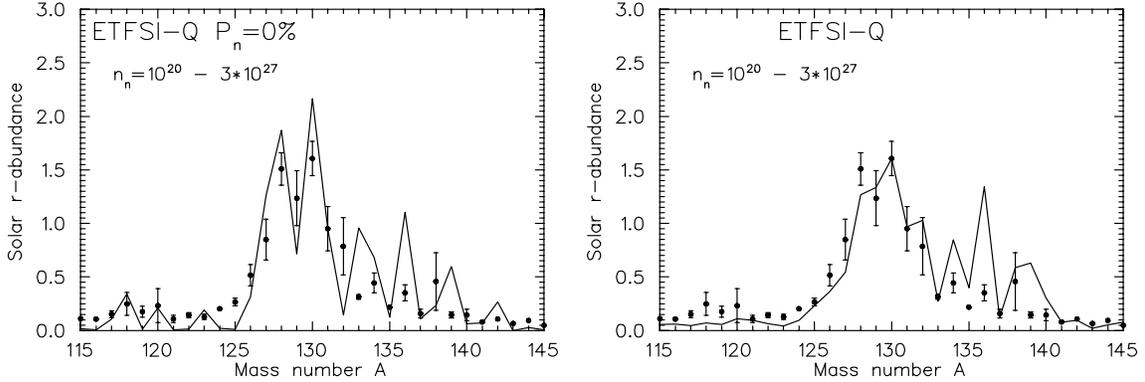,width=15.cm,angle=90}}
\caption{R-abundances in the A$\simeq$130 region from dynamical calculations 
with S$_n$ values from the ETFSI-Q mass model. In the left part, the initial 
progenitor abundances are shown, prior to $\beta$-decay back to stability; 
in the right part, the final abundances of the stable isobars are depicted.
}
\label{klk4}
\end{figure}

Since we want to focus here on the N=82 waiting-point region, we present in 
Figs.~\ref{klk3},\ref{klk4}, a cut-out in linear scale of our latest 
$\beta$-flow r-abundance calculations. They are plotted together with the 
latest N$_{r,\odot}$ ''residuals'' from the compilation of Arlandini 
et al. \cite{klk:arl}. From a comparison of these figures, 
one can draw several conclusions which may clarify the situation about the 
build-up of the A$\simeq$130 N$_{r,\odot}$ peak and the time needed to 
overcome this bottle-neck towards heavier r-isotopes. In Fig.~\ref{klk3}, we 
first show two ''fits'' with the initial simplistic assumption of constant 
S$_n$=2~MeV values for all N=83 isotopes between $_{40}$Zr and $_{49}$In, 
in one case ''normalized'' to the top of the N$_{r,\odot}$ peak at A=130 
(left part, where N=82 $^{130}$Cd is the main progenitor) and in the other 
case to the bottom of the left wing of the peak at A=125 (right part, where 
N=82 $^{125}$Tc is the main progenitor). One can easily imagine that both 
presentations may give rise to completely misleading interpretations. The 
assumption of a direct T$_{1/2}$--N$_{r,\odot}$ correlation, would -- on the 
one hand -- imply that {\bf only} $^{130}$Cd, $^{131}$In and the major part 
of $^{129}$Ag would be synthesized in $\beta$-flow equilibrium (see left side 
of Fig.~\ref{klk3}); on the other hand, all N=82 isotopes between $^{124}$Mo 
up to $^{128}$Pd would be produced in a steady flow, but not $^{129}$Ag to 
$^{131}$In (see right part of Fig.~\ref{klk3}). At this point, I cannot 
refrain to point out, that this latter picture is the basis for the recent 
conclusions of Refs. \cite{klk:mp,klk:lp}. These authors argue that {\it 
''the time spent at the N=82 waiting point is shorter than previously 
assumed''} 
%because 
and {\it ''$^{129}$Ag and $^{130}$Cd cannot be produced in 
$\beta$-flow equilibrium''}. As demonstrated above, this conclusion is at 
least carried too far, if not wrong. Therefore, it is by all means unjustified 
to further conclude that their above result would imply {\it ''that the N=82 
and N=126 r-process peaks are made at two distinct sites''}. 
In any case, if the above authors were right, their conclusions would 
detract our contribution to the success in measuring r-process waiting-point
nuclei, simply because other nuclear-physics quantities, such as 
neutron-capture cross sections, would substitute for the astrophysical 
significance of the $\beta$-decay properties in equilibrium scenarios.

In Fig.~\ref{klk4}, we now present the results obtained with S$_n$ values 
from the ETFSI-Q mass model, for the N=83 isotones varying between 0.01~MeV 
for $_{40}$Zr via 1.09~MeV for $_{44}$Ru to 2.22~MeV for $_{48}$Cd. For 
further comparison, the left part shows the initial r-abundances of the 
waiting-point nuclei prior to $\beta$-decay back to stability; and in the 
right part the final r-abundances of the stable isobars after back-decay are 
depicted. It is evident already from the first glance, that with this 
approach the N$_{r,\odot}$ pattern of the whole N=82 waiting-point region 
between A=125 and A=133 is much better reproduced than under the simplified 
conditions leading to the two fits shown in Fig.~\ref{klk3}. Since exactly 
the same $\beta$-decay quantities (T$_{1/2}$ and P$_n$) and astrophysics 
parameters (stellar temperature, neutron density, flow-time and weighting of 
the individual r-components) have been used, the improvements are clearly due 
to the increasing S$_n$ values of the (N$_{mag}$+1) isotopes on the rising 
part of the A$\simeq$130 peak. Those N=82 waiting-point isotopes at the 
bottom of the peak -- with correspondingly ''low'' S$_n$(N=83) values -- 
collect r-abundance fractions for a much larger neutron-density range than 
the ones at the top of the peak with ''high'' S$_n$(N=83) values. To be more 
specific, while in the $_{46}$Pd isotopic chain N=82 $^{126}$Pd is the major 
waiting-point nucleus up to n$_n$$\simeq$10$^{24}$~cm$^{-3}$, the magic shell 
in the Ag chain is already passed at n$_n$$\simeq$10$^{23}$~cm$^{-3}$ and in 
the Cd chain even as early as n$_n$$\simeq$10$^{22}$~cm$^{-3}$. Around 
n$_n$$\simeq$10$^{23}$~cm$^{-3}$, the actual waiting-point in the Cd isotopic 
chain is 96~ms N=84 $^{132}$Cd \cite{klk:Han00}. Around 
n$_n$$\simeq$10$^{24}$~cm$^{-3}$, even N=86 $^{134}$Cd takes over, 
and in the Ag chain N=84 $^{131}$Ag becomes the waiting-point nucleus. 
Hence, the final picture shown in the right part of Fig.~\ref{klk4} represents 
the sum of several (here altogether 16) partial equilibria, each for a 
definite neutron density, respectively r-process path. One may extend this 
discussion also to the regions below and beyond the actual abundance peak, 
where still some deficiencies exist. When focussing on the A=134 to A=140 
area, which just has become within experimental reach /cite[klk:she], the 
rather pronounced odd-even deviations seem to indicate an earlier onset of 
collectivity in very neutron-rich $_{50}$Sn to $_{52}$Te isotopes than 
commonly predicted. This would result in slightly higher S$_n$ values in 
this region, thus shifting the r-process path further away from stability 
involving progenitor isotopes with shorter T$_{1/2}$.
 
Finally to the question, why to present the {\it initial progenitor abundances} 
in the left part of Fig.~\ref{klk4}, which are no direct ''observables''? 
The reason is that in several recent papers discussing nucleosynthesis in the 
high-entropy bubble scenario of a SN~II, charged-current neutrino reactions 
are predicted to significantly substitute for $\beta$-decays. This would
require an extension from the classical $\beta$-flow equilibrium to a 
''weak-flow'' equilibrium (see, e.g. \cite{mcl}). Also in this context, it 
was stated that a weak steady-flow equilibrium at N=82 cannot be attained, 
which again would exclude $^{129}$Ag and $^{130}$Cd from this equilibrium. 
Qian et al. \cite{klk:QIAN} argue that such neutrino-processing of the initial 
r-abundances during freeze-out might help shorten the time-scale for the 
r-process and could be responsible for the ''filling-up'' of the low-mass 
wings of the A$\simeq$130 and A$\simeq$195 N$_{r,\odot}$ peaks. However, in 
this paper effects from $\beta$dn-emission during decay back to 
stability have been ignored completely. Based on our results shown in 
Fig.~\ref{klk4}, where the left part corresponds to the {\it ''unprocessed 
distribution''} of Fig.~2 in \cite{klk:QIAN}, we come to different 
conclusions. Already without considering additional abundance changes from 
capture of remaining seed neutrons and re-capture of delayed neutrons during 
freeze-out, we estimate that neutrino postprocessing effects are about an 
order of magnitude smaller than postulated by Qian et al. \cite{klk:QIAN}. 
Our calculations suggest that the main effect of ''filling-up'' the initial 
under-abundances in the A=124--126 region (see left part of Fig.~\ref{klk4}) 
is due to $\beta$dn-branching during the first 150 ms of the freeze-out 
rather than neutrino-induced neutron spallation.

\section{Summary}

In this short review, I have tried to summarize our progress in studying 
nuclear properties of very neutron-rich isotopes that may be of importance 
in r-process nucleosynthesis. Although it has been a long and painstaking 
way over more than 15 years, there are still many open experimental and 
theoretical questions, among them also just the point discussed above about 
the astrophysical relevance of our data. If a careful evaluation of detailed 
network calculations would come to the conclusion that the nuclear-physics 
properties of the ''classical'' N=82 waiting-point nuclei at the top of the 
A$\simeq$130 peak were, indeed, unimportant, we would accept it -- most 
reluctantly. However, as long as this is not proven, we will endorse with 
Claus Rolf's recent affirmationion at CGS-10: {\it ''We will measure (it) 
anyhow...!''}.

\section*{Acknowledgements}

I am very grateful to all my collaborators in the various nuclear-structure
and astrophysics studies that formed the basis for this paper. In 
particular, I would like to thank Peter M\"oller and Friedel Thielemann 
for their patience in solving -- or in terms of a nuclear-chemist 
{\it dissolving} -- many problems. This work was supported 
by various grants from the German BMBF, DFG and FCI.

\end{document}